\documentclass[10pt]{iopart}

\usepackage{iopams,graphicx}

\begin{document}

\title{Stochastic Stokes origami: folds, cusps and skyrmionic facets in random polarisation fields}

\author{Kerr Maxwell and Mark R Dennis}

\address{EPSRC Centre for Doctoral Training in Topological Design \\ and School of Physics and Astronomy, \\ University of Birmingham, Birmingham B15 2TT, UK}
\ead{m.r.dennis@bham.ac.uk}

\begin{abstract}
We consider the jacobian of a random transverse polarisation field, from the transverse plane to the Poincar\'e sphere, as a Skyrme density partially covering the sphere.
Connected domains of the plane where the jacobian has the same sign---patches---map to facets subtending some general solid angle on the Poincar\'e sphere.
As a generic continuous mapping between surfaces, we interpret the polarisation pattern on the sphere in terms of fold lines (corresponding to the crease lines between neighbouring patches) and cusp points (where fold lines meet).
We perform a basic statistical analysis of the properties of the patches and facets, including a brief discussion of the percolation properties of the jacobian domains.
Connections with abstract origami manifolds are briefly considered. 
This analysis combines previous studies of structured skyrmionic polarisation patterns with random polarisation patterns, suggesting a particle-like interpretation of random patches as polarisation skyrmionic anyons.
\end{abstract}

\vspace{2pc}
\noindent{\it Keywords}: Poincar\'e sphere, random field, jacobian, topological charge, continuous mapping, Whitney cusp, origami manifold.

\ioptwocol

\section{Introduction}\label{sec:int}

Optical beams with transverse, position-dependent polarisation demonstrate a wide range of geometric, topological and singular features \cite{roadmappol}.
For instance, such a 2D polarisation pattern displays \emph{polarisation singularities}, actively studied for over 40 years \cite{nyecline,nyehajnal,nyebook,bd01,berrycrimea,freundpoincare,mrd2002,fmsam2002,amms2002,fodp:pol,nanopolsings,dop:pio,ambzz:polsings}, notably C points of right- or left-handed circular polarisation, at which the polarisation azimuth and vibration phase are singular, and L lines of linear polarisation on which the polarisation ellipse handedness is not defined.
Their occurrence in \emph{random polarisation fields}---modelling polarisation speckle \cite{brosseau,goodman:speckle}---creates a 2D stochastic geometry which has received particular interest\cite{freundpoincare,mrd2002,fmsam2002,amms2002,fodp:pol,ambzz:polsings}.

More recently, the study of patterns of varying polarisation has shifted from the study of special points to topological polarisation \emph{textures} over 2D areas.
In particular, full Poincar\'e beams and skyrmionic beams, which realise all possible states of elliptic polarisation, are envisaged as mappings of the transverse plane wrapping around the Poincar\'e sphere, possibly multiple times  \cite{fullpoincare,skyrmionicbeams,sugic:hopfion,topquasi}.
Such textures require careful optical design, involving nongeneric structures such as loops in the transverse plane on which the elliptic polarisation is fixed, point-compactifying the loop interior, a topological disk, in order to map surjectively onto the Poincar\'e sphere.

We aim here to investigate the topological structure of random polarisation patterns, viewed as partial skyrmionic textures. 
To do this, we generate random 2D transverse polarisation fields: a polarisation ellipse---and Stokes parameters---at each $x,y$ point \cite{brosseau}.
The net polarisation pattern induces a continuous \emph{random Stokes map} $\mathcal{S}$ from the physical $\mathbb{R}^2$ plane to the Poincar\'e $S^2$ sphere.

A typical continuous mapping between two topological spaces exhibits a \emph{singular structure} pattern.
For a mapping between two surfaces (2-dimensional manifolds), this singular structure is well known since the seminal work of Whitney to consist of \emph{folds} and \emph{cusps} \cite{whitney,porteous}.
For mappings between higher-dimensional spaces, these singularities are described by higher-order catastrophes \cite{postonstewart}.
(This appearance of catastrophe theory is distinct from its role in describing optical caustics in ray patterns \cite{nyebook,bu:co,wavesnrays}.)

\begin{figure}
    \centering
    \includegraphics[width=0.8\linewidth]{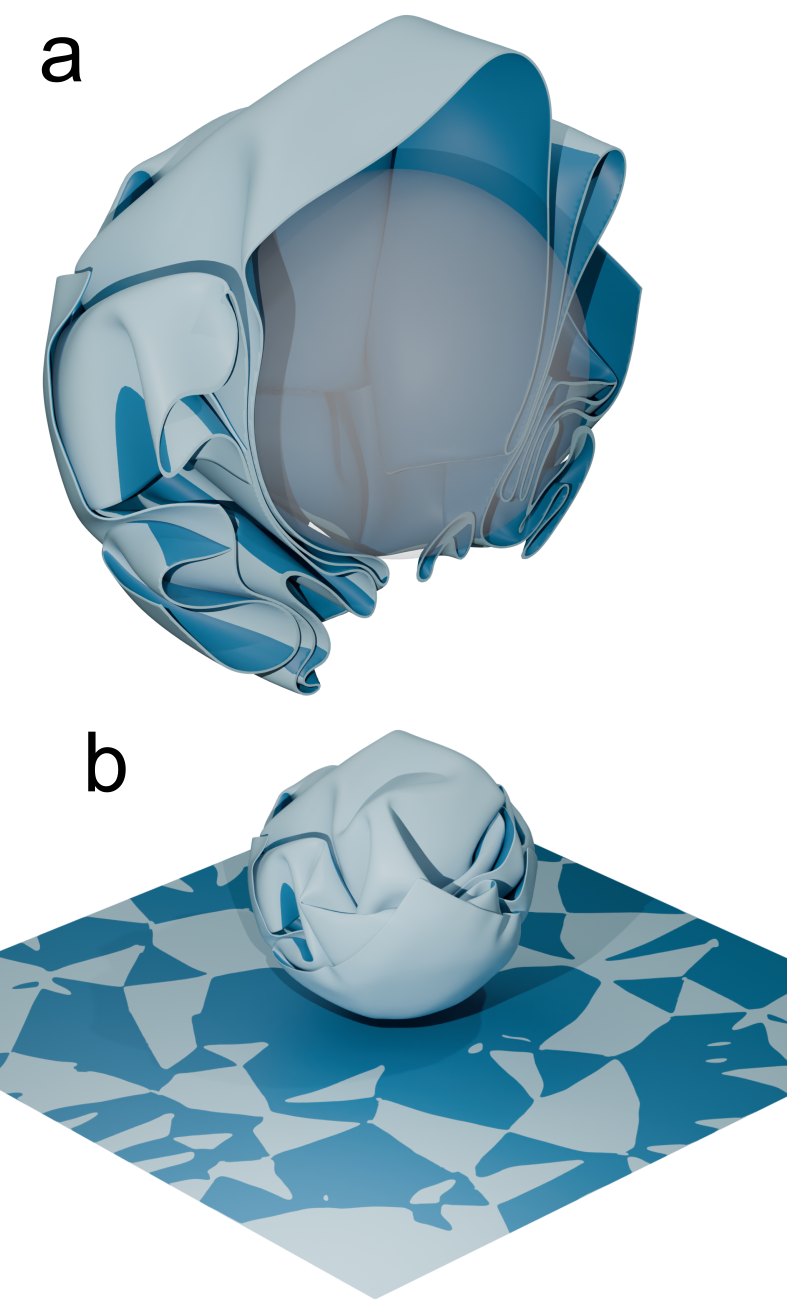}
    \caption{Random crumpling of a planar sheet wrapped around a sphere.
    (a) Cutaway of a 2-sided sheet randomly wrapping around a sphere forming a complicated folding pattern.
    The sheet colour changes at the fold lines: light blue when the upper side faces outwards, and dark blue when the lower side faces outwards.
    (b) When the sheet is flattened out, it forms a network of light and dark blue patches, meeting along crease lines (corresponding to fold lines).  
    }
    \label{fig:crumpling}
\end{figure}

Consider taking a flat piece of paper and crushing it around a sphere, as in Figure \ref{fig:crumpling}.
The paper folds over itself and sometimes forms pleats as it wraps around the sphere, possibly many times. 
We take the analogy of origami paper folding, where upon unwrapping the origami paper and flattening it out, lines folded into the paper open up to become creases.
We thus refer to the lines in the plane that map to folds on the sphere as \emph{crease lines}.
Domains of the plane bordered by crease lines are called \emph{patches}, and their images on the sphere, bounded by folds, are called \emph{facets}.
We will colour facets/patches where the upper side of the sheet points outwards (positive) from the sphere are light blue, and where they point inwards (negative) are dark blue.
Across a fold, the paper doubles back to cover the same part of the sphere, but with the side (colour) reversed, so neighbouring patches on the plane are coloured differently.
On the sphere, pairs of fold lines, bounding a pleat, meet at \emph{cusp points}.
They unfold to \emph{pleat points}. 
All of these features occur in the topological interpretation of stochastic polarisation fields.

Figure \ref{fig:crumpling} provides a visualisation of a random Stokes mapping from the transverse plane (the paper) to the Poincar\'e sphere parametrising the elliptic polarisation state (see \cite{brosseau}).
This Stokes mapping $\mathcal{S}:\mathbb{R}^2 \longrightarrow S^2$ is determined by the normalised \emph{Stokes vector} $\bi{S}(x,y) = (S_1,S_2,S_3)$ at each point of the plane labelled by $x,y$, where $S_j, j=1,2,3$ are position-dependent Stokes parameters, defined below.
The topological and singular structures of the Stokes map are determined by the behaviour of the \emph{jacobian} $\rho = \rho(x,y)$ of the Stokes map $\mathcal{S}$.
We will see that the crease lines correspond to the nodal lines $\rho(x,y) = 0$, implying the domains of connected $\mathrm{sign}(\rho)$ correspond to the patches: positive where $\rho > 0$, and negative where $\rho < 0$.
Sometimes pairs of fold lines meet at a point with antiparallel tangents, at a cusp.
We refer to the points on crease lines mapping to these cusps as \emph{pleat points}

We will study these structures (patches/facets, creases/folds and pleat points/cusps) in the stochastic polarisation pattern based on the distribution, over the transverse plane, of the signed scalar $\rho$.
Thus the Skyrme-topological interpretation of optical polarisation fields manifests even in random light beams, suggesting new interpretations of the polarisation field structure, including the polarisation singularities.
In general, the C points and L lines are distinct from the fold and cusp singularities of the Stokes map.

Our approach can be summarised in terms of mathematical singularity theory and point-set topology as follows.
Around each point in the transverse plane, consider a small, simply-connected area---a \emph{neighbourhood}---and the elliptic polarisations of the points in the neighbourhood.
This defines a small family of polarisations, corresponding to a small area on the Poincar\'e sphere (i.e.~solid angle), smoothly parametrising the states of elliptic polarisation. 
At a generic point, the Stokes map from the plane neighbourhood to the Poincar\'e sphere has rank 2: the plane neighbourhood looks like a stretched, squeezed or sheared deformation of the corresponding area on the Poincar\'e sphere (if $\rho > 0$) or such a pattern where the pattern has undergone a reflection (if $\rho < 0$).
For instance, the images of C points are the Poincar\'e sphere's poles; the nearby solid angles can be considered as regular lemon C points \cite{nyebook,berrycrimea,freundpoincare,mrd2002}.
This neighbourhood on the Poincar\'e sphere, for a skyrmionic polarisation pattern, contributes to the so-called ``polarity number'' $p$, ``vorticity number'' $m$ via the azimuthal angle \cite{topquasi}.

However, with codimension 1, the mapping rank at the point is unity (on crease lines/folds), and with codimension 2, the rank is zero (at pleat points/cusps).
In this sense, the current paper is an analogous description to $\bi{\Omega} = 0$ critical lines in 3D scalar fields, which organise the vortex topology structures \cite{bd:events}.
We will describe the effect of these structures on the geometry and topology in generic polarisation fields, explicitly via their behaviour in polarisation speckle.

The paper will have an exploratory structure, and is structured as follows.
In the next section, we describe the standard geometric and topological properties of random polarisation fields and what this means topologically for the Stokes map on the Poincar\'e sphere.
This leads to the definition of the jacobian $\rho$ in section \ref{sec:rho}, which determines the patches and creases in the transverse plane, and their images on the Poincar\'e sphere, the facets and folds.
A short analysis of their statistical distributions follows in section \ref{sec:stats}, followed by a discussion of the abstract formulation of origami manifolds in section \ref{sec:origami}, concluding with a more general discussion.

\section{Random wave fields in the transverse plane and on the Poincar\'e sphere}\label{se:random}

We denote a point in the transverse plane by $\bi{r} = (x,y)$, perpendicular to propagation in $z$.
A monochromatic, transverse $\bi{E}$-field is a complex 2-component vector field in this plane, 
 \begin{equation} 
    \bi{E}(x,y) \; = \; \sum_{\bi{k}} \left[ a_{\bi{k}} \bi{e}_+ + b_{\bi{k}} \bi{e}_- \right] \exp(\rmi (\bi{k}\cdot\bi{r})),
    \label{eq:Edef}
 \end{equation}
expressed as a superposition of plane waves with transverse wavenumbers $\bi{k}$ and complex scalar coefficients $a_{\bi{k}}$, $b_{\bi{k}}$ weighting the right- and left-handed circular polarisation vectors $\bi{e}_{\pm}=\frac{1}{\sqrt{2}}(1,\pm \rmi)$.
The ellipse is traced out by the real vector $\mathrm{Re}[\bi{E}\rme^{-\rmi \omega t}]$ as $t$ varies over $2\pi/\omega$.

\begin{figure*}
    \centering
    \includegraphics[width=0.8\linewidth]{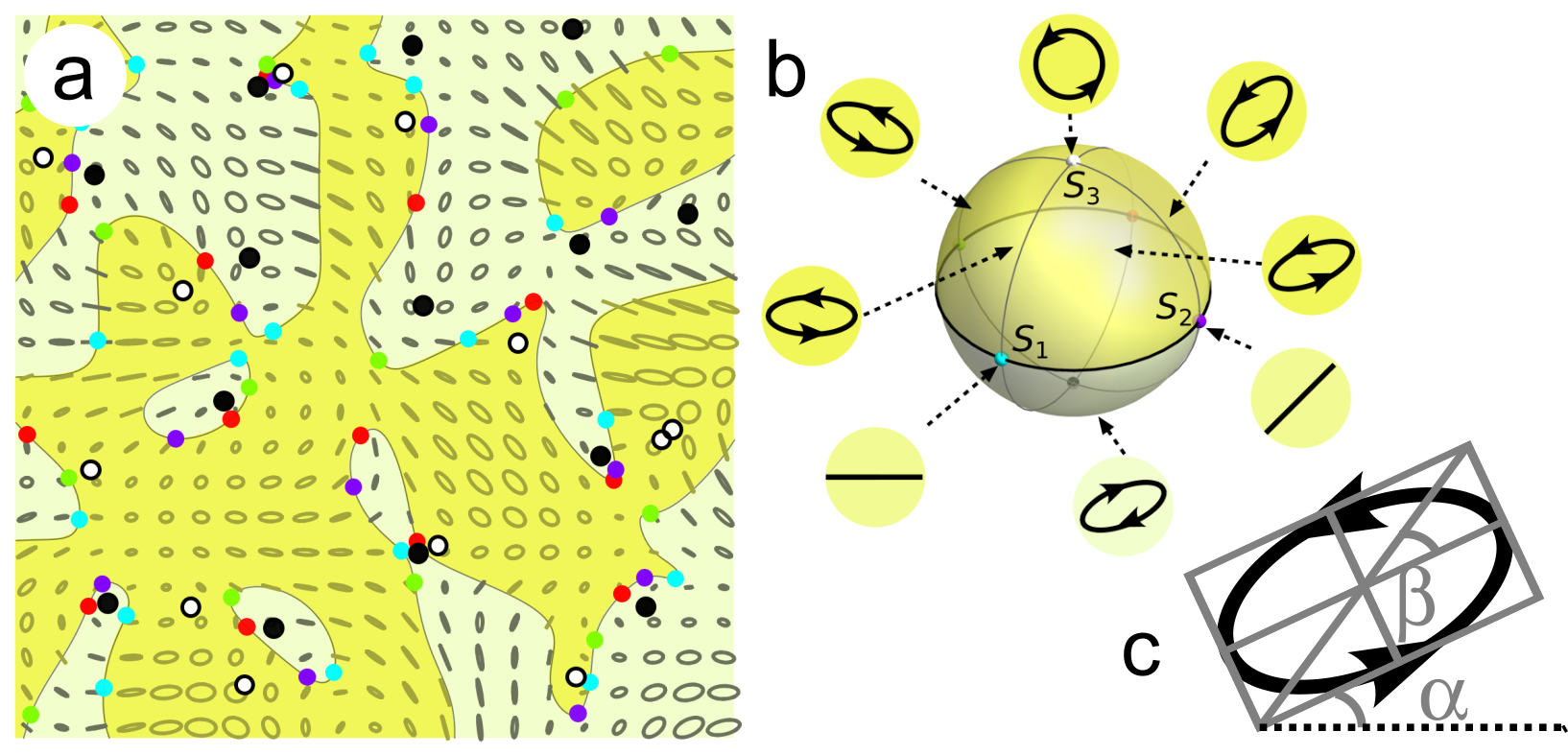}
    \caption{
    Random polarisation field.
    (a) shows $4\lambda \times 4\lambda$ of the isotropic random polarisation field described in the text around Eq.~(\ref{eq:Edef}), with fixed transverse wavenumber $k = 2\pi/\lambda$, and $\lambda$ the transverse wavelength.
    There is a polarisation ellipse at each $\bi{r}=(x,y)$ point in the transverse plane, with size proportional to total intensity $S_0$.
    Regions of RH elliptic polarisation (yellow) are separated from LH polarisation (cream) by L lines (black), with points of circular polarisation, circular RH $\bi{S}=(0,0,1)$ (white dots) and LH $\bi{S}=(0,0,-1)$ (black dots).
    The other Stokes axes occur for linear polarisations, horizontal/vertical $\bi{S}=(\pm 1,0,0)$ (cyan/red), and $45^{\circ}/135^{\circ}$ $\bi{S}=(0,\pm 1,0)$ (green/purple).
    (b) shows the Poincar\'e sphere, with RH (LH) polarisations in the north (south) hemisphere, with sphere azimuth $\phi = 2\alpha$, the ellipse azimuth.
    Coloured dots again show the Stokes axes.
    (c) shows a typical polarisation ellipse, together with its parametrising angles $\alpha, \beta$.}
    \label{fig:randomfield}
\end{figure*}

For a \emph{Gaussian random field}, a very large number of transverse wavevectors $\bi{k}$ are sampled with uniformly random directions. 
The complex coefficients are sampled from independent identically complex Gaussian random variables, ensuring the random phases are uniformly distributed.
Figure \ref{fig:randomfield} (a) shows an example random polarisation field, corresponding to a superposition of 50 random transverse wavevectors chosen with the same magnitude $k$.

The Poincar\'e sphere parametrises the possible states of polarisation, independent of overall intensity and phase.
At a fixed point $\bi{r}$, the polarisation state $\bi{E}$ is described by the three normalised Stokes parameters $S_1, S_2, S_3$, corresponding to the constants of 2D elliptic harmonic motion, for $j=1,2,3$ taken cyclically,
\begin{equation}
   S_j \; = \; \frac{\bi{E}^*\cdot\sigma_{j-1}\bi{E}}{\bi{E}^*\cdot\bi{E}},
   \label{eq:stokesparas}
\end{equation}
where $\sigma_1, \sigma_2, \sigma_3$ are the three Pauli matrices. 
At slight variance with this notation, we write $S_0 = \bi{E}^*\cdot\bi{E}$ as the total intensity.
The normalisation ensures the Stokes parameters satisfy $S_1^2+S_2^2+S_3^2 = 1$, and hence the space parametrising all polarisation states is a sphere -- the \emph{Poincar\'e sphere}. 
The Poincar\'e sphere is represented in Figure \ref{fig:randomfield} (b).

Topologically, the position-dependent polarisation pattern is the continuous Stokes mapping $\mathcal{S}$ from the plane to the sphere, as discussed above.
This map is not globally invertible in general: many different points $\bi{r}$ may correspond to the same polarisation state $\bi{S}$. 
This is equivalent to stating that each particular Stokes vector $\bi{S}'$ on the sphere may have many \emph{preimages} $\bi{r}_j$, $j=1,2,\ldots$ in the plane, where for each $j$, $\bi{S}(\bi{j}) = \bi{S}'$.
For instance, all right-handed C points are preimages of the north pole $\bi{S} = (0,0,1)$.
Figure \ref{fig:randomfield} (a) shows the preimages of the six axes in Stokes space $\bi{S} = (\pm 1,0,0), (0,\pm 1,0), (0,0,\pm 1)$ (sometimes called Stokes vortices \cite{freundpoincare,mrd2002,fmsam2002}) in the random sample function.

The Stokes vector encodes the natural geometric parameters of elliptic polarisation: the polarisation \emph{azimuth angle} $\alpha$, $0 \le \alpha < \pi$, subtended by the ellipse major axis with the $x$-axis, and the \emph{ellipticity angle} $\beta$, $-\pi/4 \le \beta \le \pi/4$, whose tangent is the signed ratio of ellipse minor/major axes, as represented in Figure \ref{fig:randomfield} (c).
These are closely related to cylindrical coordinates on the Poincar\'e sphere: azimuth $\phi$ and height $Z$:  
\begin{equation}
   \phi \; \equiv \; \arg(S_1 + \rmi S_2) = 2\alpha, \quad   Z \; \equiv \; S_3 = \sin2\beta.
   \label{eq:PScyl} 
\end{equation}
The C points are those where $Z = \pm 1$ (right handed/left handed), i.e.~$\beta = \pm \pi/4$, and the azimuth $\phi,\alpha$ at those points is not defined.
The L lines in the plane are those where $Z = 0$, i.e.~$\beta = 0$, and the orientation of linear polarisation can have any $\alpha$.

\section{The jacobian $\rho$ of the Stokes map}\label{sec:rho}

\subsection{Poincar\'e patches and facets}

\begin{figure*}
    \centering
    \includegraphics[width=0.8\linewidth]{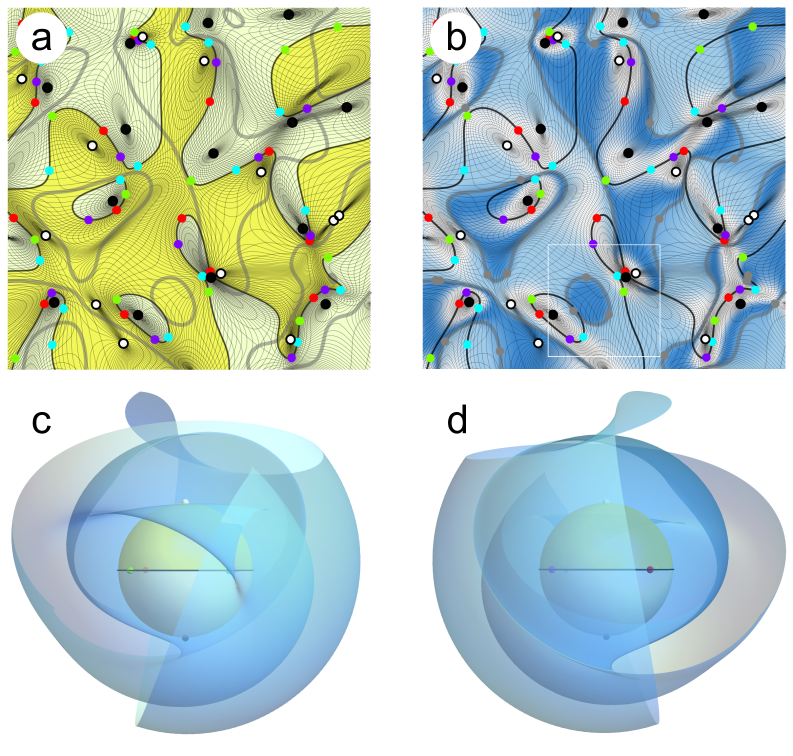}
    \caption{Creases, patches and facets of the random polarisation field.
    (a) shows the same plot as Figure \ref{fig:randomfield} (a), but in terms of contours of azimuth $\phi=2\alpha$ and ellipticity $Z = S_3=\cos\beta$.
    These pick out the C points as circulations of $\phi$ and maxima/minima of $Z$.
    The crease lines $\rho(x,y) = 0$ (grey curves) are loci where the two sets of contour lines are (anti)parallel.
    (b) shows the same plot as (a), now in terms of $\rho$, picking out positive ($\rho > 0$, light blue) and negative ($\rho < 0$ dark blue) patches whose boundaries are the crease lines ($\rho = 0$, grey). 
    They are decorated by pleat point (grey points).
    Some patches contain areas of high topological charge $|\rho|$ (white).
    (c), (d) show different views of the sphere, with the mapping from the white square in (b) partially unwrapped, showing folds and cusps.}
    \label{fig:creasefold}
\end{figure*}

Via the Stokes map, we define functions in the physical plane $\phi(x,y) = 2\alpha(x,y)$ and $Z(x,y)$, representing azimuth and height on the Poincar\'e sphere (i.e.~ellipse azimuth and ellipticity).
For our random sample function, their contours are shown in Figure \ref{fig:creasefold} (a): in particular, C points occur at the centre of small loops of $|Z| \lesssim 1$, which are phase singularities where all lines of constant $\phi$ meet.
Thus every point $\bi{r}$, except for the C points, is surrounded by a small rectangle bounded by contours of $\phi$ and $Z$, with area proportional to $\nabla \phi \times \nabla Z \cdot\bi{e}_z$, i.e.~proportional to the cross product of the gradients of the ellipse parameters $\alpha$ and $\beta$.
This corresponds to the area form in the transverse plane
\begin{equation}
   (\nabla \phi \times \nabla Z \cdot\bi{e}_z) \rmd x \wedge \rmd y \; = \; \rmd \phi \wedge \rmd Z.
   \label{eq:2formrho}
\end{equation}
The rectangle area is therefore the \emph{jacobian} $\rho$ of the Stokes map,
\begin{equation}
   \rho \; = \; \nabla \phi \times \nabla Z \cdot\bi{e}_Z \; = \; \partial_x \phi \partial_y Z - \partial_y \phi \partial_x Z,
   \label{eq:rhodef}
\end{equation}
determining the rate at which solid angle area of the Poincar\'e sphere is covered with respect to area in the plane.
This is a measure of the magnitude of the 2-dimensional polarisation state change with 2-dimensional position generalising the vector gradient of a single function.

On the sphere, $\phi$ increases eastwards and $Z$ increases northwards.
Thus $\rmd \phi \wedge \rmd Z$ is positively oriented on the Poincar\'e sphere, and the corresponding weighting from 3D Stokes space to the surface of the sphere is unity, i.e.~$\bi{S}\cdot\partial_{\phi}\bi{S}\times\partial_Z\bi{S} = 1$.

In the plane, the sign of $\rho$ is determined by the sign of the cross product of $\nabla \phi= 2\nabla \alpha$ with $\nabla Z$.
$\nabla \alpha$, perpendicular to the contours of constant $\alpha(x,y) =\phi(x,y)/2$, is the direction of maximum azimuth change in the polarisation ellipse field, and $\nabla Z$, perpendicular to contours of $Z(x,y) = S_3(x,y)$, is the direction of maximum change of signed ellipticity.
In the plane, $\nabla Z$ might point to the right of $\nabla \alpha$ ($\rho > 0$), or to to the left ($\rho <0$), and for Gaussian random fields, each occurs with equal probability.

The jacobian function $\rho$ is of fundamental importance to topological and skyrmion physics.
In a full skyrmion within the plane, its integral over the skyrmion area is $4\pi$ times an integer.
It is, therefore, sometimes referred to as the continuous \emph{topological charge density}, or \emph{Skyrme density}, as it corresponds to the rate that the sphere solid angle is covered, with respect to a 2D skyrmion's area in the plane.
Its analogue in topological photonics is called the \emph{Berry curvature} \cite{topphotonics}, integrating to an integer Chern number, and accumulating a net geometric phase on a path around the boundary. 
In the original setting of 3D skyrmions as models of particles in high-energy physics \cite{mantonsutcliffe}, the analogue of $\rho$ is called the \emph{baryon density}.

The sign of the jacobian $\rho(x,y)$ is the same everywhere in the same domain of the plane, i.e.~a \emph{patch} as defined in section \ref{sec:int}, with the image of a patch, on the Poincar\'e sphere, called a \emph{facet}.
Consider a point $\bi{r}'$ in a patch, with $\bi{S}'=\bi{S}(\bi{r}')$.
A small neighbourhood of $\bi{r}'$ maps to a corresponding neighbourhood around $\bi{S}'$, and a right-handed circuit of $\bi{r}'$ in this neighbourhood maps to a right-handed ($+$, orientation-preserving) or left-handed ($-$, orientation-reversing) loop on the sphere depending on $\mathrm{sign}(\rho)$.

A familiar representation of $\rho$ is in terms of $\bi{S}$,
\begin{equation}
   \rho \; = \; \rho \, \bi{S}\cdot \partial_{\phi}\bi{S}\times\partial_Z \bi{S} \; = \; \bi{S}\cdot \partial_x \bi{S} \times \partial_y \bi{S},
   \label{eq:rho2}
\end{equation}
similar to the previous sphere orientation expression (equal to $1$) but with coordinates on the plane $x,y$ replacing Poincar\'e sphere coordinates $\phi, Z$.

The integral of $\rho$ over the area of a patch corresponds to the solid angle subtended by the corresponding facet.
For full Poincar\'e beams, the whole plane corresponds to a single patch, covering the sphere once.
More complicated multiskyrmion beams and skyrmion lattices cover the sphere an integer number of times (i.e.~the Skyrme number).
As discussed in Section \ref{sec:int}, the decomposition of a simple polarisation skyrmion can be formalised into a disk or annulus on the Poincar\'e sphere, varying over $0 \le \phi \le 2\pi$ and a range of $\beta$, with the total facet area given by the total solid angle subtended on the sphere.

Unlike a designed skyrmion, the solid angle subtended by a facet in a random field can take any real value, positive or negative according to the sign of $\rho$.
In a random field, the areas these patches and facets cover can be large and small; we will consider their probability distribution later.

We consider first the ways the patches cover the plane and the facets cover the Poincar\'e sphere.
The Stokes map from patch to facet is not necessarily 1-to-1: a single facet might cover a neighbourhood of the sphere multiple times; the 
significance of the patch and facet is that the orientation sense is the same over the connected area.

The sign of the jacobian endows C points (where $S_3 = \pm 1$) with the sign of their topological $\pm 1/2$ index \cite{nyebook,berrycrimea,mrd:monstardom}.
The image of a small circuit around a C point encloses either the corresponding north or south pole of the Poincar\'e sphere.
This must pass through all values of $\alpha$, from $0$ to $\pi$, with either the same orientation $+1/2$ or opposite $-1/2$.
The precise classification (lemon, star, monstar) of the C point morphology depends on further structure \cite{mrd:monstardom}.
Similarly, along an L line oriented with the right-handed region to the left (i.e.~$\nabla Z$ points to the left), the azimuth angle $\alpha$ increases for $\rho > 0$, decreases for $\rho < 0$, according to the sign of $\nabla \alpha \times \nabla Z \cdot \bi{e}_z$.
This also determines the signs of the ``Stokes vortices'' where $S_1 = \pm1$ and $S_2 = \pm 1$ \cite{mrd2002,fmsam2002,amms2002}.

The magnitude of $\rho$ determines the rate area on the sphere is covered per unit area in the plane.
At a fixed point $\bi{r}$, $1/\rho(\bi{r})$ represents the ratio of $\rmd x \rmd y / \rmd \phi(\bi{r}) \rmd Z(\bi{r})$: the area of an infinitesimal rectangle of $x$ and $y$ contours around $\bi{r}$, and $\rmd \phi(\bi{r}) \rmd Z(\bi{r})$, corresponding to an infinitesimal area on the sphere around $\bi{S}(\bi{r})$.
When $\rho$ is large, a large sphere area is covered by a small area on the plane (many $\phi,Z$ parallelograms with respect to plane area in Figure \ref{fig:creasefold} (a)), whereas when $\rho$ is small, only a small sphere area is covered (few $\phi,Z$ parallelograms with respect to plane area in Figure \ref{fig:creasefold} (a)).

A map of the patches for our sample random function is shown in Figure \ref{fig:creasefold} (b).
Some of the patches are finite and enclosed entirely in the plotted map; other large patches extend beyond the represented area and may be finite or infinite.
A striking feature of several patches is regions where total intensity $S_0$ is very small, corresponding to large values of $\rho$; much of the net solid angle covered by the facets correspnds to these areas.
These areas of super-large $\rho$ are analogues of regions of random superoscillation, as is well-studied in speckle patterns \cite{dhc:superosc,roadmapsuperosc}.
We will consider these large statistical fluctuations more systematically below in section \ref{sec:stats}.

The Stokes map itself for a part of the random field example is shown in Figure \ref{fig:creasefold} (c), (d): in addition to folding, and pleating near cusps, the plane is stretched and squeezed around the sphere according to the magnitude of $|\rho|$.

\subsection{Creases and folds}

Consider how an area neighbourhood of a point on a crease line behaves under $\mathcal{S}$. 
The image on the sphere is folded over, reversing its natural orientation on one side and so changing the sign of the jacobian $\rho$ upon crossing the boundary between patches. 
This is to say, locally, the map covers only only one side fold line.
These features can be seen in Figure \ref{fig:creasefold} (c), (d).
Topologically, this shows the presence of folds in the image shows that $\mathcal{S}$ is not surjective, and hence not invertible.
\footnote{$\mathcal{S}$ is nevertheless continuous as a neighbourhood of a point on the fold is mapped from (i.e.~has preimage of) two---or more---distinct neighbourhoods in the plane.}.

In the plane, the fold preimages are the \emph{crease lines}, following the paper mapping analogy, as discussed in section \ref{sec:int}.
The unfolding topology implies that polarisation state at each point near a crease line must also occur near the other side of the crease line, but with the orientation of polarisations in a neighbourhood reversed.
The crease lines constitute the boundaries of the network of patches in the plane---domains of $\mathrm{sign}(\rho)$---forming the singular structure pattern (Figure \ref{fig:creasefold} (b)), similar to the domains of ellipse handedness $\mathrm{sign}(Z)$ bordered by $L$ lines (Figure \ref{fig:creasefold} (a)).

More generally, crease lines can be expressed in a coordinate-free way as the loci where the rank of the Stokes mapping becomes unity, rather than two as it is generically.
However, our discussion will focus on their properties in terms of $\alpha(x,y), Z(x,y)$.

Since $\rho$ is proportional to $\nabla \alpha \times \nabla Z$, the crease lines $\rho = 0$ are characterised as the points where the gradients of $\alpha$ and $Z$ are (anti-)parallel.
Thus along crease lines, the contour lines of $\alpha(x,y)$ and $Z(x,y)$ are parallel, and the rectangle swept out between the $\alpha, Z$ contours collapses to a degenerate line, perpendicular to $\nabla \alpha$ and $\nabla Z$.

The functions $\alpha(x,y)$ and $Z(x,y)$ may have saddle points.
As singular points of direction with $-1$ Poincar\'e index, saddle points neutralise the net positive, discrete topological charge acquired at the C points: circulations of $\nabla \alpha$ and sources/sinks of $\nabla Z$ \cite{mrd2002,dop:pio}.
At azimuth saddle points, $\nabla \alpha = 0$, with all vector directions nearby; this agrees with $\nabla Z$ at the saddle point in some direction, thus a crease line passes through all saddle points of $\alpha$. 
A similar argument applies to saddle points of $Z$.

For a typical random function, as shown in Figure \ref{fig:creasefold} (a), (b), the crease lines are smooth curves in the plane, closing as loops enclosing a patch of finite area, or percolating over the whole plane (bounding an infinite, percolating patch), in a similar way to zero contours of other sign-symmetric functions on the field, such as the L lines bounding domains of handedness $\mathrm{sign}(Z) = \mathrm{sign}(S_3)$.
They do not generically cross other crease lines, except at specific events when an external parameter is varied (codimension 3).
We will discuss percolation further in section \ref{sec:stats}.
At such points, the crossings evidently occur at loci where saddles in $\alpha$ and in $Z$ coincide and align.

\subsection{Cusps}

According to catastrophe theory, there are two generic singular local geometries arising in continuous 2D maps, such as $\mathcal{S}$; fold lines (discussed above), and \textit{cusp} points. 
In the image space, cusps appear where two fold lines meet with the same tangent vector, at the beginning of a pleat in the folded surface.
Viewing these two branches as a single, continuous fold line, cusps are points on the continuous fold lines where the tangent vector reverses direction.
At these points, the rank of the Stokes map is zero.

As defined in Section \ref{sec:int}, we will refer to the preimages of cusps as \emph{pleat points} in the plane, lying on crease lines.
Being points of zero rank, $\nabla \phi$ and $\nabla Z$ are perpendicular to the crease line tangent at the pleat points.
This may be seen as follows: an infinitesimal line element $\rmd\sigma(s) = \sqrt{[\rmd\phi(s)]^2 + [\rmd Z(s)]^2}$ along a fold line, is parametrised by arc length $\rmd s  = \sqrt{\rmd \bi{r}\cdot \rmd \bi{r}}$ of its preimage crease line in the plane.
We define $\chi$ as the angle between $\rmd \sigma$ and the east-pointing $d\phi$, so on the fold line, $\tan\chi = \rmd Z/\rmd \phi$.
Since 
\begin{eqnarray}
   \rmd \phi & = & \partial_x \phi \rmd x + \partial_y \phi \rmd y \; = \; \nabla \phi \cdot \rmd \bi{r}, \nonumber \\
   \rmd Z & = & \partial_x Z \rmd x + \partial_y Z \rmd y \; = \; \nabla Z \cdot \rmd \bi{r}, \label{eq:metrics}
\end{eqnarray}
the line element on the fold, in terms of $\rmd s$ on the crease line (on which $\nabla \phi$ and $\nabla Z$ are (anti)parallel), is
\begin{eqnarray}
   \rmd\sigma(s) & = & \sqrt{[\nabla\phi\cdot \rmd \bi{r}]^2 + [\nabla Z\cdot \rmd \bi{r}]^2} \nonumber \\
   & = & \rmd s (\nabla \phi \cdot \bi{t}) |\sec\chi|,
   \label{eq:dsig}
\end{eqnarray}
where $\bi{t} = \rmd \bi{r}/\rmd s$, the unit tangent vector on the crease line, and for simplicity we assume $\chi \neq \pm \pi/2$.
Thus at a point $s = s_0$ on the crease line at which $\nabla \phi \cdot \bi{t} = 0$, then $\rmd \sigma = 0$: this is the pleat point.
Furthermore, as $s$ crosses $s_0$, $\nabla \phi \cdot \bi{t}$ changes sign, reversing $\rmd \sigma/\rmd s$: the tangent on the fold line reverses at the cusp point.

At the pleat points, the jacobian $\rho = 0$ and, since a line perpendicular to a curve's tangent is (anti)parallel to its normal,
\begin{equation}
   \nabla \rho \times \nabla \phi \cdot \bi{e}_z \; = \; \partial_x \rho \partial_y \phi - \partial_y \rho \partial_x \phi \; = \; 0.
   \label{eq:cuspcond}
\end{equation}
In the rare event that $\nabla \phi = 0$ at the cusp point where $\nabla Z \cdot \bi{t} = 0$ (i.e.~the cusp tangent make an angle $\chi = \pm \pi/2$), then these expressions can be rearranged in terms of $\nabla Z$ instead of $\nabla \phi$.
As can be seen in Figure \ref{fig:creasefold} (c), (d), there are typically many cusps on the folds in the image of $\mathcal{S}$, and hence many pleat points on crease lines in Figure \ref{fig:creasefold} (b).

\subsection{Local model of cusp, fold and crease}

\begin{figure}
    \centering
    \includegraphics[width=0.8\linewidth]{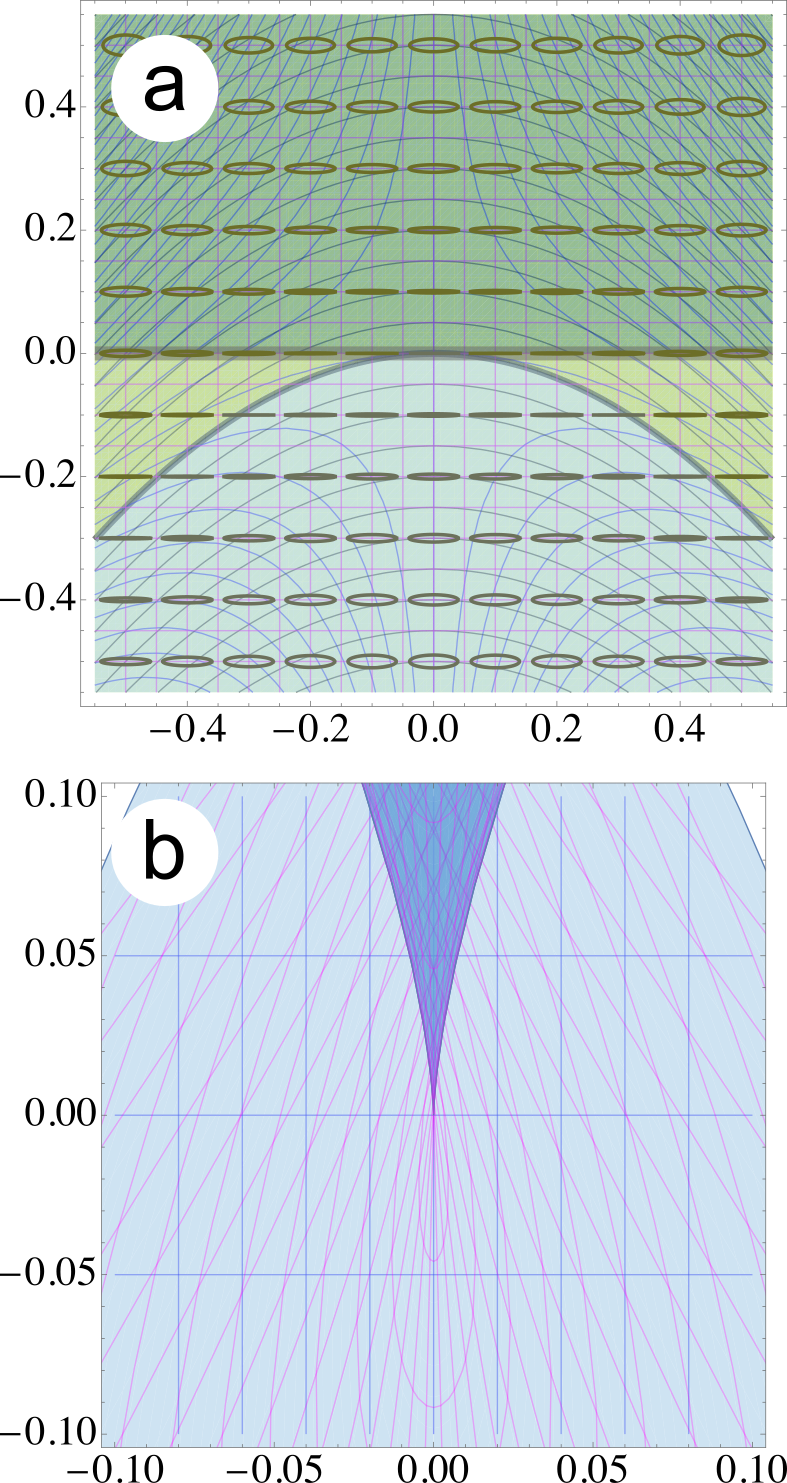}
    \caption{Model cusp field of Equation (\ref{eq:model}). 
    (a) shows the polarisation ellipse field, the contours of $\alpha$ and $Z$ (blue), the L line and crease line, and a system of $x,y$ grid lines (pink).
    The pleat point is at the origin.
    (b) shows the $(S_2,S_3)$ plane tangent to $\bi{S} = (1,0,0)$, the polarisation state at the origin, together with the images of the $x,y$ grid (pink), and the $\alpha, Z$ grid (blue).}
    \label{fig:cusp}
\end{figure}

In order to visualise the structure of a crease, mapping to a fold decorated by a cusp, we consider a local model of a Stokes map (expanded to third order),
\begin{equation}
   \bi{S}_{\mathrm{mod}}(x,y) = (1 - \frac{1}{2} y^2 - x^2 y, -x y - \frac{2}{3} x^3, y + x^2 ).
   \label{eq:model}
\end{equation}
This field is shown in Figure \ref{fig:cusp} (a), and is equivalent to the standard cusp normal form \cite{porteous,postonstewart}.
At the origin, $\bi{S}_{\mathrm{mod}}(0,0) = (1,0,0)$; the polarisation is horizontal linear, and in fact there is an L line through the origin corresponding to $S_{3\mathrm{mod}} = 0$, $y = -x^2$.
$S_2$ has a saddle point at the origin.

From the Stokes map, the field gradients (appropriately expanded) can be calculated:
\begin{eqnarray}
   \nabla \alpha & = & \frac{1}{2}\left(-y - 2 x^2 - \frac{1}{2}y^3, x + \frac{3}{2} x y^2\right), \label{eq:gradalpha} \\
   \nabla Z & = & \left( 2x , 1 \right) \label{eq:gradZ}.
\end{eqnarray}
The vector direction for $\nabla Z$ rotates with small angle $-2x$, from the direction $(0,1)$ when $x = 0$.
$\nabla \alpha$ has a saddle point at $x = 0$, and along $y = 0$, is $\nabla \alpha \approx (-x^2,x/2) = (x/2)\nabla Z$.
Thus along $y = 0$, for $x > 0$, $\nabla \alpha$ and $\nabla Z$ are parallel, and antiparallel for $x < 0$.
This is a crease line with a pleat point at $(0,0)$: we expect it to map to a fold line with the origin mapping to a cusp.

As it maps to the neighbourhood of the Stokes axis $(1,0,0)$, on the Poincar\'e sphere, the model field can be considered in the tangent plane $(S_2,S_3)$ only.
This is shown in Figure \ref{fig:cusp} (b).
The image of the crease line is the parametric curve $(S_2,S_3) = (-2x^3/3,x^2)$, corresponding to two fold lines following the curves $S_2 = \pm 2 S_3^{3/2}/3$, meeting, at $(S_2,S_3) = (0,0)$ when $x = 0$, the cusp/pleat point.

At a point near and below the crease line $(x,-\delta y)$, for $x>0$, $\delta y \ll x$, the Stokes vector is 
\begin{eqnarray}
   \bi{S}_{\mathrm{mod}}(x,-\delta y) & \approx & (1 - x^2 \delta y, x \delta y - \frac{2}{3} x^3, -\delta y + x^2 ) \nonumber \\
   & \approx & \bi{S}_{\mathrm{mod}}(x-\delta y/x,\delta y),
   \label{eq:modelmirror}
\end{eqnarray}
so each polarisation state below and near the crease line corresponds to a polarisation state above the crease line, justifying the image as a fold.

\section{Basic statistical analysis of $\rho$, patches, and facets in random polarisation fields}\label{sec:stats}

In the language of this paper, the main interest in polarisation skyrmions is having finite or infinite patches of the transverse plane mapping to cover the Poincar\'e sphere an integer number of times.
However, we have seen that in random polarisation patterns, there is no simple global statement we can make about the covering from the whole plane to the sphere (except, on average, it is zero).
Nevertheless, individual random patches cover the Poincar\'e sphere as facets of net positive or negative solid angle, and arbitrarily small patches have been found to occur.
What is the area $A_p$ of a given patch $p$ in the transverse plane? 
What is the area/solid angle $SA_p$ of the corresponding facet?
These give the spatial average of $|\rho|$ over patch $p$, $\overline{|\rho|}_p$, as $SA_p/A_p$.

We will see below in the random wave model (\ref{eq:Edef}) that the global average of the jacobian modulus, $\langle |\rho| \rangle$, is simply $K_2$, the second moment of the probability distribution of transverse wavenumbers \cite{mrd2002,lh,bd:321,mrd2007}.
In the example of the plots, with is a fixed transverse wavenumber $k$, this second moment $K_2 = k^2$ \cite{mrd2002,bd:321,mrd2007}, so the average topological charge density is one steradian per squared inverse wavenumber.
The distribution of the patches in the sample field of Figure \ref{fig:creasefold} (b) shows spatially extended patches (extending beyond the plotted area) of both signs, with smaller, finite-area patches embedded.

\begin{figure}
    \centering
    \includegraphics[width=0.8\linewidth]{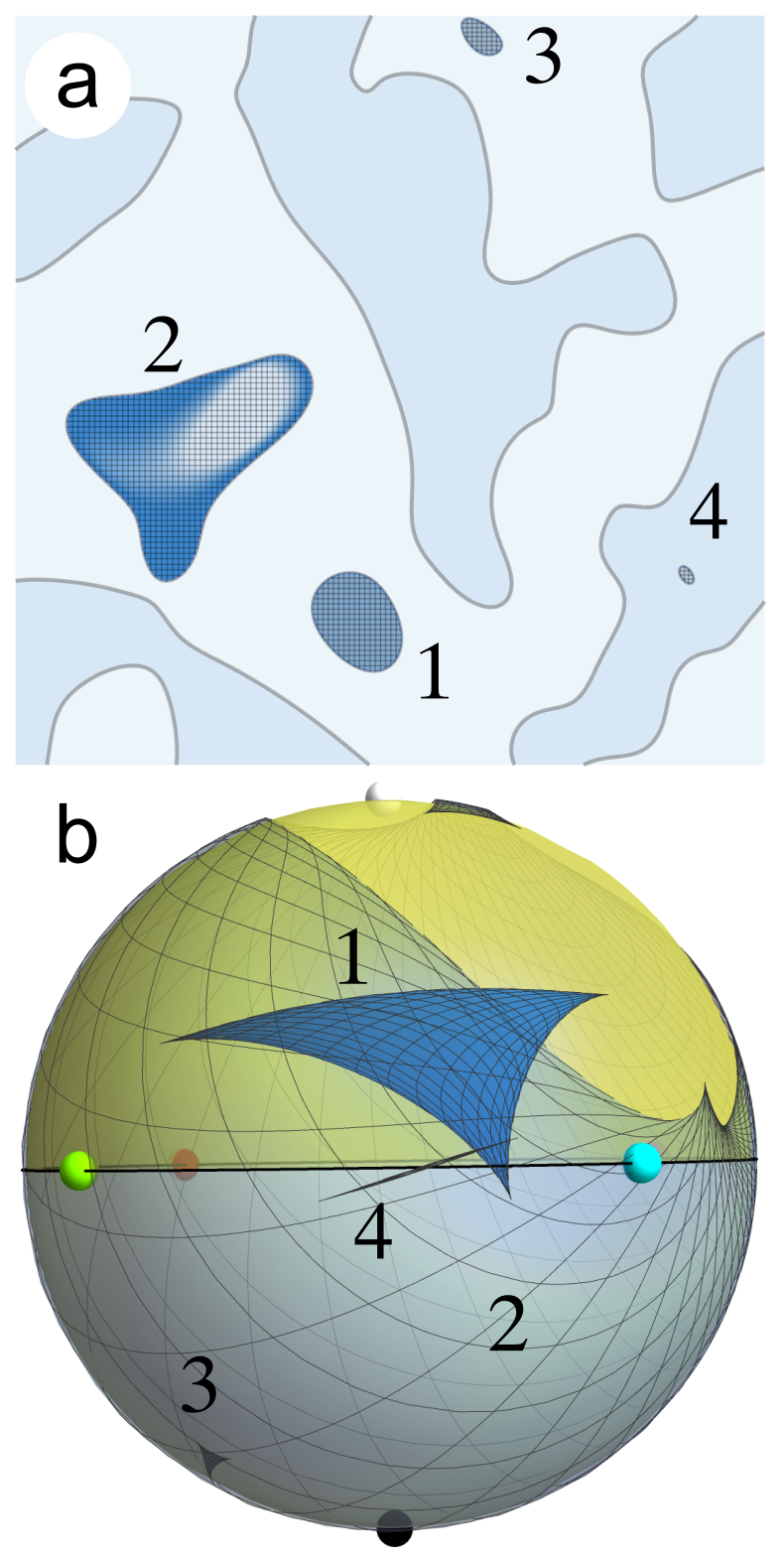}
    \caption{Finite area patches and corresponding facets.
    (a) shows the previously plotted random field area, with patches 1,2,3,4 highlighted, patches 1,2,3 are negative, and patch 4 is positive.
    (b) shows the corresponding facets, with facet 2 much larger than the others, covering a significant fraction of the sphere.
    The areas of these patches and facets are given and discussed in the main text.}
    \label{fig:domainfacet}
\end{figure}

Figure \ref{fig:domainfacet} breaks this down directly: in the transverse plane of Figure \ref{fig:domainfacet} (a), within the plotted area of 16 square wavelengths (i.e.~$64 \pi^2 k^{-2}$), there are four finite patches: patches 1,2,3 where $\rho < 0$ (patch 2 being somewhat larger than the others), and, much smaller, patch 4 where $\rho > 0$.
These map to facets on the sphere in Figure \ref{fig:domainfacet} (b), with facet 1 clearly having 3 cusps, facet 2, with 3 cusps, covers a large fraction of the sphere, facet 3 is very small but just has 3 visible cusps, facet 4, overlapping facet 1, is very small and elongated.
Numerical integration finds the patches have areas $A_1 = 1.91 k^{-2}$, $A_2 = 8.18 k^{-2}$, $A_3 = 0.28k^{-2}$, $A_4 = 0.36k^{-2}$.
The corresponding facet areas are $SA_1 = 0.200$, $SA_2 = 11.082$, $SA_3 = 0.007$, $SA_4 = 0.020$, each in steradians.

\

Averaging over each patch, $\overline{|\rho|}_1 = 0.105k^2$, $\overline{|\rho|}_2 = 1.354 k^2$, $\overline{|\rho|}_3 = 0.024k^2$, $\overline{|\rho|}_4 = 0.056 k^2$.
Therefore, patches 1,3,4 have a significantly smaller average topological charge density of $k^2$, and patch 2 is slightly larger than average.
This can be understood in terms of the regions of low total intensity and rapid polarisation change, represented in Figures \ref{fig:creasefold} (b) and Figure \ref{fig:domainfacet} (a) as white regions, and discussed above as a polarisation analogue of random superoscillations \cite{dhc:superosc}.
The topological charge density $\rho$ is much larger than average in these regions; there is one such in patch 4, but none in the other 3 patches.
This is consistent with the spatial averages $\overline{|\rho|}_{1,2,3,4}$ above.

The random field defined in (\ref{eq:Edef}) is a Gaussian random variable at each point; it is statistically stationary (averages are invariant to translations and rotations), and is ergodic (spatial averages $\overline{\bullet}$ correspond to ensemble averages $\langle \bullet \rangle$).
Many statistical properties of these fields may be calculated analytically, such as the density of RH/LH C points as $K_2/4\pi \approx 0.0796 K_2$ \cite{mrd2002,goodman:speckle}; since the ensemble is statistically rotation symmetric, this is the density of points of any particular fixed state of polarisation (i.e.~point on the Poincar\'e sphere), such as the preimages of the Stokes axes (Stokes vortices) as coloured dots in Figure \ref{fig:creasefold}.

The density of L lines is $\sqrt{K_2}\pi/4\sqrt{2} \approx 0.555 \sqrt{K_2}$ \cite{mrd2002}.
This is somewhat higher than the density of real nodal lines (of say $\mathrm{Re}E_x$), which we discuss below, is is $\sqrt{K_2}/2\sqrt{2} \approx 0.22 \sqrt{K_2}$ \cite{mrd2007,berry:lines}.

Direct application of the methods in \cite{mrd2002,bd:321,mrd2007}, and analogous to the superoscillation calculation of \cite{dhc:superosc}, it is straightforward to show that the probability density of $\rho$ itself, $P(\rho)$, over the ensemble of random functions is given by:
\begin{equation}
   P(\rho) \; = \; K_2^{-1} \left(1 + |\rho|/K_2\right)^{-3},
   \label{eq:rhoPDF}
\end{equation}
in terms of the second moment of the transverse wavenumber spectrum $K_2$.
Clearly, this is sign-symmetric, $\langle \rho \rangle = 0$ (the jacobian is as likely to be negative as positive), and, as stated above, $\langle |\rho|\rangle = K_2$.
The distribution has slowly-decaying algebraic tails: moments higher than the first diverge.
This reflects the fact, discussed above, that the rate of change of polarisation with position is much greater in regions where the normalising intensity $S_0$ is small; this is a 2D sphere analogue of the natural, random superoscillations of phase in random scalar wavefields \cite{dhc:superosc}.
In particular, as an example, the fraction of the area where $|\rho| > K_2$ is $75\%$ of the plane, and where $|\rho| > 4K_2$ is $36\%$ if the plane.

As $\rho$ is a complicated cubic combination of fields and their derivatives, it is not easy to calculate more complicated quantities associated with $\rho$ ; unlike C points and L lines, it appears hard, for example, to calculate analytically either the density of crease lines or the pleat points.
From different realisations of random fields with $K_2 = k^2$, we estimate the density of crease lines to be $(0.5659 \pm 0.0006)k$, which is $1.019 \pm 0.001$ of the L line density; thus these two line densities are numerically close.
To estimate the plane density of pleat points, we numerically locate the nodal points of the complex scalars $\rho + \rmi(\bi{e}_x\cdot\nabla\rho \times \nabla \alpha)$ and $\rho + \rmi(\bi{e}_x\cdot\nabla\rho \times \nabla Z)$.
These are the points on crease lines where the contours of $\rho$ align with the contours of $\alpha$ and of $Z$, including the saddle points; with these saddle points, either of these sets of nodal points overcounts the pleat points, but their intersection should agree with their true count.
We find this density to be $(0.90 \pm 0.04)k^2$ which is $1.79\pm 0.08$ times the density of any point of fixed polarisation (including LH and RH C points).
This error is quite large, and we observe this may underestimate the pleat point density: we find the nodal points used to locate the cusp points are highly anisotropic and are thus hard to locate using standard numerical vortex algorithms \cite{methodology}.

This crease line density is numerically quite close to the L line density, and, given the error, the pleat point density may in fact be twice the fixed polarisation point density (i.e., equal to the total C point density).
This closeness has precedents in the numerical closeness between statistical L line densities and zeros of the Gaussian curvature of random functions and the density of C points in transverse fields with C$^T$ and L$^T$ points in non-transverse random fields, despite different analytic forms.
Significantly more numerical effort would be required to understand this better for crease lines and pleat points.

\begin{figure*}
    \centering
    \includegraphics[width=0.8\linewidth]{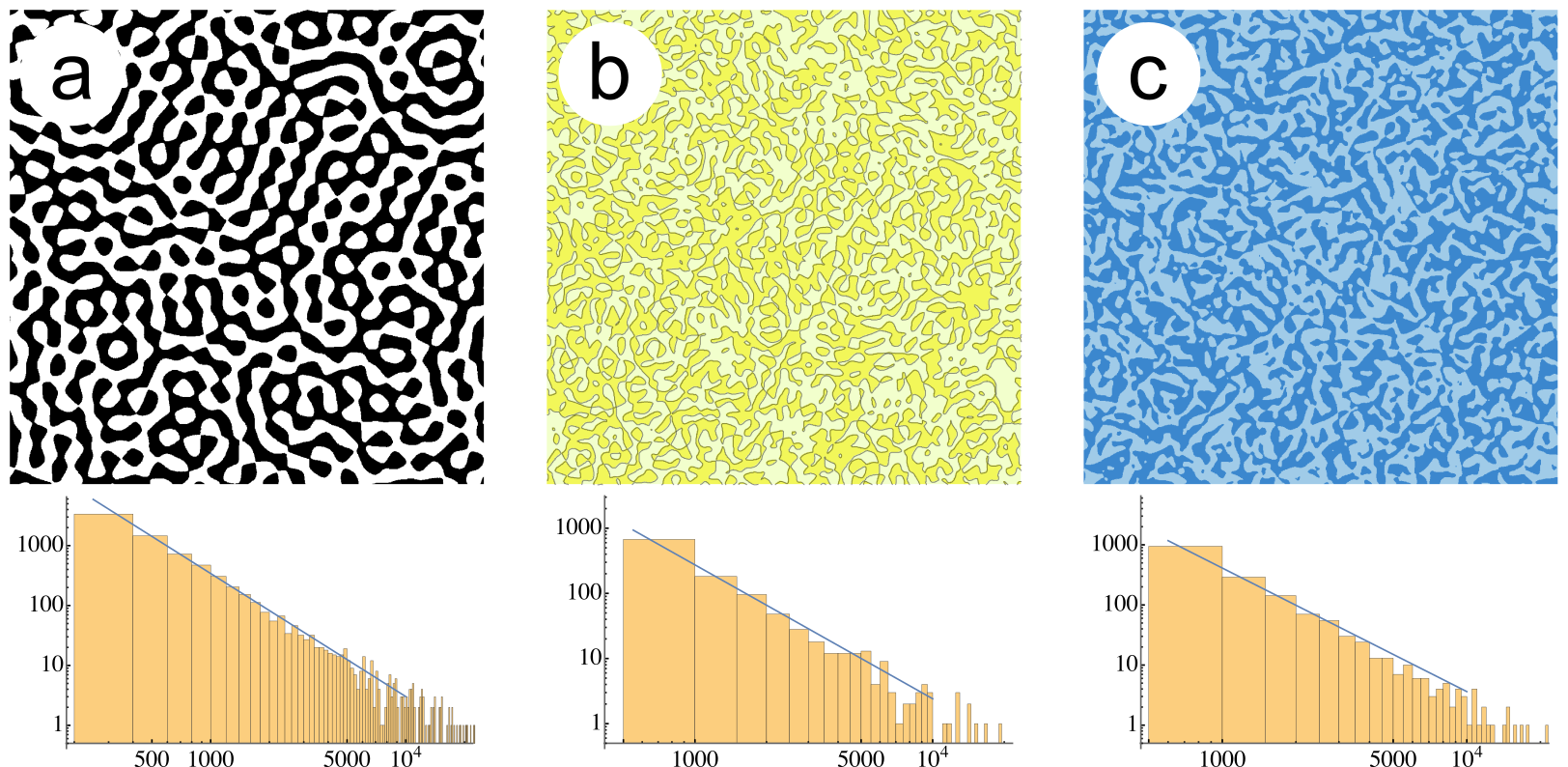}
    \caption{Large-scale areas of the same random field, demonstrating percolating domains.
    In each case, the positive-negative domains of some sign-symmetric function $f(x,y)$ associated with the random polarisation field are plotted over $400\lambda^2$.
    In each case the random domains of each sign take a range of different sizes, including percolating domains that cross the whole area.
    In addition each case has a log-log histogram of domains by area, and fitted to a straight line of $-187/91$ given by percolation theory.
    (a) Nodal domains of the real wavefield $\mathrm{sign}( \mathrm{Re}[E_x])$; (b) Right-/Left-handed domains $\mathrm{sign}(S_3) = \mathrm{sign}(Z)$, separated by L lines; (c) jacobian patches $\mathrm{sign}(\rho)$, separated by crease lines.}
    \label{fig:perc}
\end{figure*}

On a larger scale, one can see across the plane that a complicated network of differently sized patches covers the plane, looking similar to the nodal domains of a random wave (taken here to be $\mathrm{Re}[E_x]$), and the R/L handedness domains separated by L lines (as shown in Figure \ref{fig:perc}).
It was proposed by Bogomolny and Schmit \cite{bs:percolation} that sign-symmetric systems of random domains like these might be in the universality class of 2D percolation, whose scaling properties are given by conformal field theory, and power law exponents are rational numbers.
In particular, they found a good agreement between the scaling of histograms of real random function domain sizes.
We show an equivalent plot here, taken over an area $10^6 k^{-2}$ of our random function in Figure \ref{fig:perc} (a).
Analogous plots of handedness and jacobian sign are shown in Figure \ref{fig:perc} (b) and (c); for each, there is a plausible agreement with the percolation scaling.
Such random skyrmionic polarisation patches are good candidates for future analysis of topological models displaying critical phenomena.

Our quick analysis here only scratches the surface of the statistical properties of skyrmionic patches in random fields and the corresponding random Stokes map on the Poincar\'e sphere.
Topologically, finite-area patches might be simply connected, as those in Figure \ref{fig:domainfacet}.
There is no topological restriction for them to be annular (containing one sub-patch), or to have multiple holes (i.e.~more than one sub-patches), although this appears statistically unlikely.
A full statistical analysis would involve calculating the patches' Euler characteristic \cite{adler}, in a similar way to the calculation for L lines and related domains \cite{mrd2007}.

Furthermore, our analysis based on the random function realised here realises only random polarisation speckle patterns with a fixed transverse wavenumber $k$, in part because of its interesting long-range properties, including percolation.
We expect different, potentially short-ranged behaviour for other random wave spectra.

\section{Stochastic Polarisation Patterns as An Origami Manifold} \label{sec:origami}

We have discussed 2D elliptic polarisation patterns as maps $\mathcal{S}:\mathbb{R}^2\rightarrow S^2$, characterised by the transformation's jacobian $\rho$.
In this context, the \emph{singular structures} of $\rho$ consist of the crease lines and pleat points. 
The image of a neighbourhood of a point in the singular structure is not a neighbourhood on the sphere, instead collapsing generically to folds (along lines) and cusps (at points).
In this section we will consider how this can be considered as a concrete example of an abstract 2D \emph{origami manifold}\cite{guillemin2000unfolding,da2011symplectic}, which we propose as a natural model of the random polarisation patterns in question.
In this context these are 2D surfaces endowed with a singular but well-behaved notion of area---called a \emph{folded area form}---from which one can define an \textit{origami template} consisting of the singular structures, which efficiently describes the topology of patches/facets, creases/folds and cusps.
We will only outline this approach here, and will employ the language and notation of mathematical topology.

As usually defined, a smooth surface (2-manifold) has an area 2-form $\omega$ which can be locally written everywhere as $\rmd x\wedge \rmd y$, i.e.~$M$ can be considered as a symplectic manifold.
When $M$ is compact, $\int_M\omega$ is the global surface area of $M$.
As discussed in previous sections, the jacobian $\rho$ of a continuous map $M \longrightarrow M'$ between such surfaces vanishes along the preimage of a fold.
We can use $\rho$ to define an augmented area form  $\omega=\rho(x,y)\,\rmd x \wedge \rmd y$.
This $\omega$ vanishes along crease lines, incorporating the folded structure into the very definition of area for $M$.
Such a surface $M$ (subject to some technicalities described below) is called an \textit{origami manifold} as it can be manipulated using abstract \textit{folding} and \textit{unfolding} operations \cite{guillemin2000unfolding} and can be described using a lower-dimensional object called an origami template. 
Cutting out the singular structures from $M$, i.e.~considering $M\backslash\rho^{-1}(0)$, leaves disconnected components (i.e.~the patches). 
Each such patch can be viewed as a Hamiltonian dynamical system encoding the polarisation texture, with the 1D origami template providing rules according to which the individual patches can be glued back together.

This is a specialisation to our particular physical system of a general manifold, for which we provide the full definition below (see \cite{da2011symplectic} for a complete account).
Equip a $2n$-dimensional manifold $M$ with a 2-form $\omega$, symplectic everywhere except on a codimension 1 submanifold $Z$ where $\omega$ vanishes transversally, and the restriction $\omega|_Z$ is of maximal rank.
The pair $(M,\omega)$ is called a \emph{folded symplectic manifold}, and becomes an \emph{origami manifold} if the null foliation integrates to oriented circle fibres over a compact base (for 2D maps, effectively, the crease lines are loops).

\begin{figure*}

    \centering

    \includegraphics[width=0.8\linewidth]{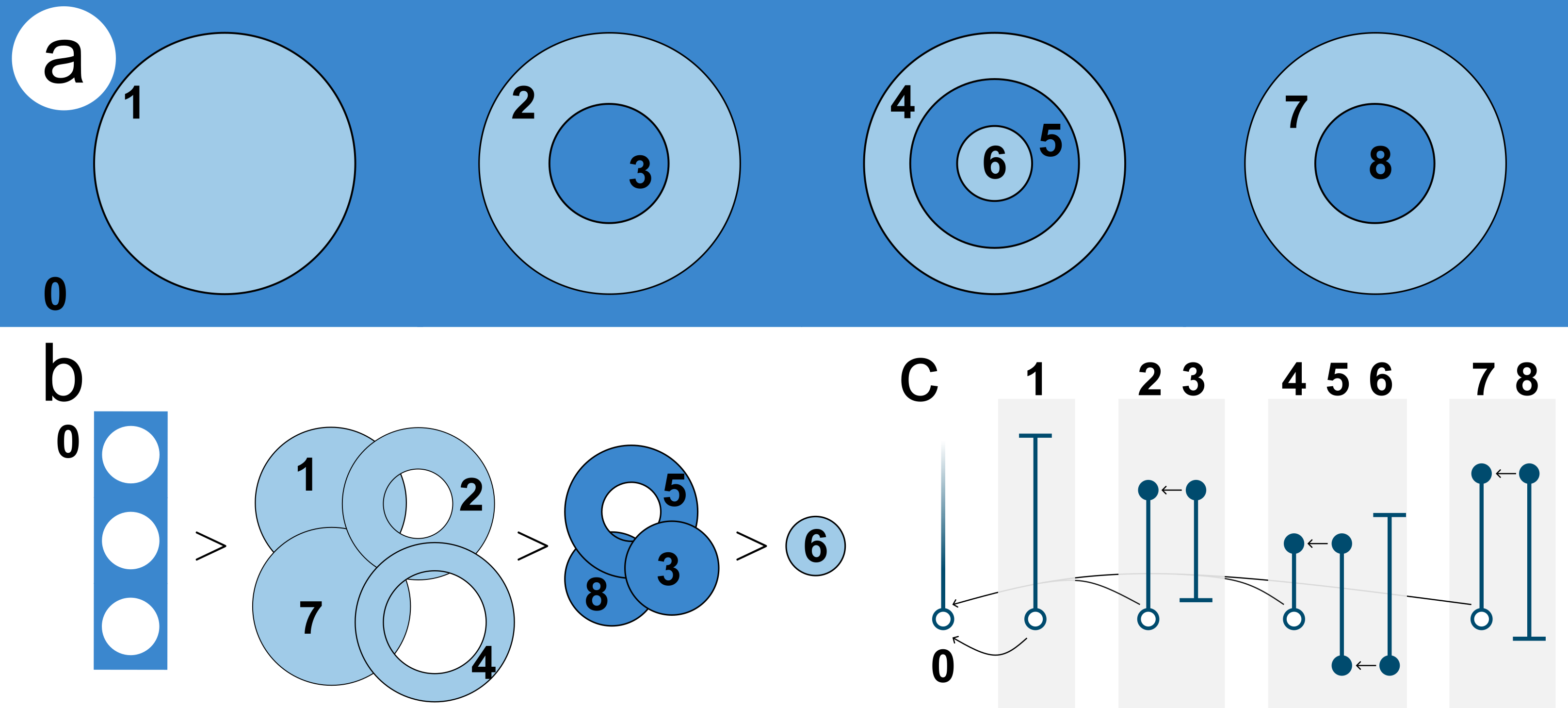}

    \caption{The origami structure of a generic Stokes map.
    (a) A generic random Stokes map consists of nested regions of the plane (topologically disks or annuli) which take on distinct values of $\mathrm{sign}(\rho)$.
    Regions of positive/negative $\mathrm{sign}(\rho)$ are coloured light/dark blue, as in previous Figures.
    We assume for simplicity that $\rho$ is stationary only at a single maximum or minimum within each disk-shaped region (1,3,6 and 8).
    (b) represents the natural partial order on the components of (a) given by the inclusions filling in the annuli.
    (c) shows the origami template of (a) viewed as an ``overlapping" of smaller templates.
    Each line represents the image of $\mathrm{abs}(\mu)$ evaluated on a particular patch, with segment length $\mu_{\mathrm{max}}-\mu_{\mathrm{min}}$.
    Two types of vertex are possible: a circular end indicating the presence of a fold (always at $\rho=0$ and a blunt end indicating a maximum/minimum of $\mu$. The black arrows represent gluing instructions. }
    \label{fig:origamiTemplate}

\end{figure*}

We now describe an example of the origami manifold and template description in action using the of a generic Stokes map, defined by patches whose boundaries are given by the zeros of $\rho$.
We consider the example with topology shown in Figure \ref{fig:origamiTemplate} (a), with finite patches a collection of topological disks and $n$-holed annuli.
These are naturally endowed with a partial ordering according from their nesting, as shown in Figure \ref{fig:origamiTemplate}(b), with the simply-connected patches, $D_i$, topologically equivalent to a disk, as minimal elements.
Each such $D_i$, equipped with $\rho_{D_i}:=\rho|_{D_i}$ (the jacobian restricted to the disk, including the boundary where $\rho$ vanishes) can be viewed as the phase space of a bound, oscillating Hamiltonian dynamical system \cite{holm2011geometric}, whose Hamiltonian $\rho_{D_i}$ generates cyclic paths about one or more extremal points $\bi{r}=\bi{r}_{\mathrm{ext}}$ where $\left| \rho_{D_i}(\bi{r}) \right|$ is a local maximum.

A famous collection of results \cite{atiyah1982convexity,guillemin1982convexity,delzant1988hamiltoniens} from the field of \textit{symplectic topology} (which includes our system as a very special case, see \cite{mcduff2017introduction} for an introduction), loosely states that the total dynamical system (as a \emph{toric} symplectic manifold, up to \emph{equivariant symplectomorphism)} is defined by the convex hull of the extremal values exhibited by the constants of the motion, which we denote by $\mu$.
We can thus redraw our space using the folded area $\rho(x,y)\,\rmd x\wedge \rmd y$ and compute the range of values takes by $\mu$, which behaves as a form of ``angular momentum" circulating in each $D_i$. Each patch is thus mapped to a line segment $\left[\mu_{\mathrm{min}},\mu_{\mathrm{max}}\right]\subseteq \mathbb{R}$, called a \emph{Delzant polytope}. 
A similar story holds for the non-simply connected patches.

The statement that $\left(\mathbb{R}^2,\rho(x,y)\,\rmd x\wedge \rmd y\right)$ is \textit{origami} implies the existence of nice gluing instructions which resolves the hierarchy of inclusions in Figure \ref{fig:origamiTemplate}(b),
This is captured, along with the dynamical information of each patch, in an \textit{origami template} \cite{pissarra2010origami}.
For $\mathbb{R}^2$, the origami template is formally an overlap of the many 1-dimensional line segments (discussed above), with lengths given the image of $\mu$ on that patch.
Patches with a common boundary are connected by a gluing instruction connecting pairs of vertices on the origami template.
The origami template for our example is given in Figure \ref{fig:origamiTemplate}(c): the overlapping components have been spread over the horizontal axis so that the gluing instructions (black arrows) and hierarchies (grey boxes) are clearly visible. Note that the line segments are glued back-to-front (instead of end-to-end) according to the requirement that the neighbourhoods of each vertex are identifiable. More work is needed to rigorously extend this analysis to patterns of patches whose nesting and critical point (extrema and saddle points)  structure is more complicated.

A major motivation for this description is the application of origami manifold analysis to 3D polarisation textures in vector beams.
This higher-dimensional case would be significantly more complicated than the given example: gluing relationships are specified between curves (not points), and there are a greater variety of singular structures to account for. However abstract methods are well suited to these higher dimensional situations.

\section{Discussion}\label{sec:disc}

We have applied the topological tools of skyrmionic fields to random polarisation fields.
We found connected domains of the jacobian sign---patches---map to facets on the Poincar\'e sphere, bounded by folds and cusps.
The jacobian---or continuous topological charge density---is strongly concentrated in regions where intensity is low and the polarisation state changes rapidly with position, the generalisation of well-known phase superoscillations in random fields to polarisation patterns.
These observations are supported by a basic statistical analysis of the random field distributions, and we identify these random maps as endowing the plane and sphere with the structure of origami manifolds.

In principle, any transverse polarisation field with position-dependent polarisation will have a structured jacobian, and will display patches, crease lines and pleat points mapping to facets, folds and cusps.
When not random, these patterns might display symmetry structures based on features of the underlying field; a skyrmionic analysis may reveal new features of such fields.
Such fields may also display structured super-polarisation gradients.
In particular, we expect the complex nanophotonic and polaritonic field structures displaying interesting polarisation singularity fields \cite{nanopolsings,bunching,polariton} to be good candidates for interesting skyrmionic facets and patches, and potentially to be analysed as origami manifolds.

Unlike a polarisation singularity analysis, which endows privileged status to circular and linear polarisation states, the skyrmionic behaviour on the sphere is completely independent of particular polarisation states and only depends on the rate at which the area is covered (i.e.~the jacobian $\rho$).
Therefore the skyrmionic analysis is invariant to any arbitrary rotation of the Poincar\'e sphere.
Furthermore, the Whitney random map applies to maps to other surfaces, such as the torus to sphere as in topological photonics \cite{topphotonics}; these patterns, too, would be decorated by creases, folds and cusps.
As loci where the rank of the map is reduced by one or two, the crease lines and pleat points constitute higher-order polarisation singularities dependent on the structure of the derivatives of the field, rather than simply points of particular states of polarisation.

The skyrmionic analysis, for any kind of structured polarisation field, could be generalised to higher dimensions, such as a map from a 3D propagating, transverse field, to the 3D optical hypersphere parametrising both phase and polarisation.
Indeed the experimental measurement of the 3D polarisation skyrmionic hopfion in \cite{sugic:hopfion} shows 3D patches where the measured $\rho$ is positive and negative.
Such 3D patches are bounded by crease surfaces where $\rho = 0$, on which are cusp lines and swallowtail points \cite{postonstewart}. 
In fact, these structures have been investigated in 3D mappings involving nonlinear skyrmions in high-energy physics, where specific mappings, arising as minimum energy configurations, have symmetries which organise the catastrophe structures of the mappings \cite{hk:folding,fk:folding}.

As discussed previously, creases and folds have an analogy in 3D scalar fields $\psi$ as the $\bi{\Omega} = 0$ lines where the scalar gradient of the field is a linear ellipse, $\nabla \psi^* \times \nabla \psi = 0$ \cite{bd:events}.
As for those loci, crease lines have the property that in a polarisation field evolving under a parameter, C points of opposite index, and therefore residing in oppositely-signed patches, will annihilate on a crease line, colliding with the special crease points where $S_0$ has a local minimum \cite{bd:events}.
Further features of crease lines may emerge from re-examination of previous literature of C point nucleation/annihilation (e.g.~~\cite{ambzz:polsings}), or by considering these structures in 3D polarisation fields.

Skyrmionic fields in linear optics display nontrivial topologies, but unlike nonlinear skyrmions in 2D and 3D condensed matter and high-energy physics, they are not usually topologically protected.
Since integer wrappings are not topologically protected in optics, the partial coverings of patches and facets studied here are, in principle, no less natural topologically than full polarisation skyrmionic fields.
Therefore, further study of partial facet coverings may deepen the connections between different types of polarisation topology in nanophotonics.

A Whitney-style analysis of patches, facets, and cusps can be applied to arbitrary surface-to-surface in a variety of physical systems.
For instance, in topological photonics, momentum space maps from the 2-torus (a square with periodic boundary conditions) to the sphere are often considered.
Such maps can be analysed in terms of patches and facets. Significantly, the usual quantity of interest in such a mapping from  compact surface to a compact surface, is the degree.
This is not only provides the Skyrme number, but also leads directly to the Chern number.
local deformations, including evolving folds and facets, cannot change the global topology.

As reflected by the literature (as reviewed by \cite{topquasi}) including other papers in this collection, as mappings displaying a quantised wrapping around the Poincar\'e sphere, standard 2D polarisation skyrmions are good examples of topological ``quasiparticles'' in optical beams.
We have seen in this work that the wrapping interpretation of facets, folds and cusps in more general polarisation patterns, despite not being quantised, has an interesting topological interpretation and relation to polarisation singularities.
In this sense, it might be productive to identify the patches while corresponding facets cover any arbitrary part of the Poincar\'e sphere, which might, as quasiparticles of structured light, be considered as optical skyrmionic anyons!

\ack
We are grateful for useful comments and suggestions from Michael Berry, J\"org G\"otte and Ricardo Heras Osorno.
Both authors acknowledge support from the EPSRC Centre for Doctoral Training in Topological Design (EP/S02297X/1).

\end{document}